\documentclass{article}
\usepackage{src/preprint}
\usepackage{afterpage}
\usepackage{algorithmic}
\usepackage{array} %
\usepackage{amsmath,amssymb,amsfonts}
\usepackage{booktabs} %
\usepackage[nocompress]{cite}
\usepackage{color}
\usepackage{graphbox}
\usepackage{graphicx}

\usepackage{mathtools}
\usepackage{multirow}
\usepackage{textcomp}

\usepackage{tikz}
\usepackage{pifont} %
\usepackage{colortbl}
\usepackage{flushend}
\usepackage[normalem]{ulem}

\useunder{\uline}{\ul}{}
\usepackage{hyperref}
\usepackage{cleveref} 
\usepackage{atbegshi} 
\usepackage{mfirstuc}
\usepackage{csquotes}

\usepackage{textcomp}
\usepackage{wrapfig}
\usepackage{relsize}
\usepackage{float}
\usepackage{placeins}
\usepackage{listings}
\usepackage{caption}
\usepackage{subcaption}

\definecolor{codebg}{gray}{0.92}
\lstdefinelanguage{ieeepython}{
    language=Python,
    morekeywords={self},
    keywordstyle=\color{black}\bfseries,
    emph={__init__},
    emphstyle=\color{black},
    stringstyle=\color{gray},
    commentstyle=\color{gray}\itshape,
    showstringspaces=false,
    basicstyle=\ttfamily\footnotesize,
    breaklines=true,
    columns=flexible,
    frame=none,
    backgroundcolor=\color{codebg},
    aboveskip=0.5em,
    belowskip=0.5em,
}

\lstdefinelanguage{ieeego}{
    morekeywords={
        break,default,func,interface,select,case,defer,go,map,struct,
        chan,else,goto,package,switch,const,fallthrough,if,range,type,
        continue,for,import,return,var,
    },
    sensitive=true,
    morecomment=[l]{//},
    morecomment=[s]{/*}{*/},
    morestring=[b]",
    morestring=[b]`,
    keywordstyle=\color{black}\bfseries,
    commentstyle=\color{gray}\itshape,
    stringstyle=\color{gray},
    showstringspaces=false,
    basicstyle=\ttfamily\footnotesize,
    breaklines=true,
    columns=flexible,
    frame=none,
    backgroundcolor=\color{codebg},
    aboveskip=0.5em,
    belowskip=0.5em,
}

\lstdefinelanguage{ieeejson}{
    morestring=[b]",
    showstringspaces=false,
    basicstyle=\ttfamily\footnotesize,
    breaklines=true,
    columns=flexible,
    frame=none,
    backgroundcolor=\color{codebg},
    literate=
     {:}{{{\color{black}:}}}{1}
     {,}{{{\color{black},}}}{1}
     {"}{{{\color{gray}"}}}{1},
    stringstyle=\color{gray},
    comment=[l]{//},
    commentstyle=\color{gray}\itshape,
    aboveskip=0.5em,
    belowskip=0.5em,
}

\newcommand{\name}{ReFaaS} %
\newcommand{\Name}{\name{}} %
\newcommand{\NameLong}{(Rewriting Functions as a Service)}

\newcommand{\anonymize}[1]{#1}

\definecolor{step}{HTML}{D80073}
\definecolor{prope}{HTML}{6A00FF}
\newcommand*\circled[1]{
    \hspace{-0.5em}
    \tikz[baseline=(char.base)]{
            \node[shape=circle,draw,inner sep=0.75pt,text=white,fill=step] (char) {#1};
    }
    \hspace{-0.5em}
}

\newcommand*\measured[1]{
    \hspace{-0.5em}
    \tikz[baseline=(char.base)]{
            \node[shape=circle,draw,inner sep=0.75pt,text=white,fill=prope] (char) {#1};
    }
    \hspace{-0.5em}
}

\newcommand{\TODO}[1]{}%

\newcommand{\orcidSW}{0000-0001-8051-7226}

\usepackage{hyperref}
\def\BibTeX{{\rm B\kern-.05em{\sc i\kern-.025em b}\kern-.08em
    T\kern-.1667em\lower.7ex\hbox{E}\kern-.125emX}}

\usepackage{tikz}

\newcommand\copyrighttext{%
  \footnotesize \textcopyright \the\year{} IEEE. Personal use of this material is permitted. Permission from IEEE must be obtained for all other uses.}

\newcommand\copyrightnotice{%
\begin{tikzpicture}[remember picture,overlay]
\node[anchor=south,yshift=10pt] at (current page.south) {\fbox{\parbox{\dimexpr0.75\textwidth-\fboxsep-\fboxrule\relax}{\copyrighttext}}};
\end{tikzpicture}%
}

\usepackage{titlesec}
\titlespacing\section{0pt}{12pt plus 3pt minus 3pt}{1pt plus 1pt minus 1pt}
\titlespacing\subsection{0pt}{10pt plus 3pt minus 3pt}{1pt plus 1pt minus 1pt}
\titlespacing\subsubsection{0pt}{8pt plus 3pt minus 3pt}{1pt plus 1pt minus 1pt}

\definecolor{lime}{HTML}{A6CE39}
\DeclareRobustCommand{\orcidicon}{
	\begin{tikzpicture}
	\draw[lime, fill=lime] (0,0)
	circle [radius=0.16]
	node[white] {{\fontfamily{qag}\selectfont \tiny ID}};
	\draw[white, fill=white] (-0.0625,0.095)
	circle [radius=0.007];
	\end{tikzpicture}
	\hspace{-2mm}
}

\usepackage{authblk}

\title{Code once, Run Green: Automated Green Code Translation in Serverless Computing}

\author{Sebastian Werner~\href{https://orcid.org/\orcidSW}{\orcidicon}}
\author{Mathis Kähler}
\author{Alireza Hakamian~\href{https://orcid.org/0000-0001-9899-0062}{\orcidicon}}
\affil{University of Hamburg, Germany \\
\texttt{\{sebastian.werner, alireza.hakamian\}@uni-hamburg.de} \\
Technische Universität Berlin, Germany \\ 
\texttt{mathis.kaehler@campus.tu-berlin.de}}  
\begin{document}

\twocolumn[\begin{@twocolumnfalse}
\maketitle

\copyrightnotice

\begin{abstract}
The rapid digitization and the increasing use of emerging technologies such as AI models have significantly contributed to the emissions of computing infrastructure.
Efforts to mitigate this impact typically focus on the infrastructure level such as powering data centers with renewable energy, or through the specific design of energy-efficient software.
However, both strategies rely on stakeholder intervention, making their adoption in legacy and already-deployed systems unlikely.
As a result, past architectural and implementation decisions continue to incur additional energy usage -- a phenomenon we refer to as energy debt.

Hence, in this paper, we investigate the potential of serverless computing platforms to automatically reduce energy debt by leveraging the unique access to function source code. 
Specifically, we explore whether large language models (LLMs) can translate serverless functions into more energy-efficient programming languages while preserving functional correctness.
To this end, we design and implement \name{} and integrate it into the Fission serverless framework.
We evaluate multiple LLMs on their ability to perform such code translations and analyze their impact on energy consumption.

Our preliminary results indicate that translated functions can reduce invocation energy by up to 70\%, achieving net energy savings after approximately 3,000 to 5,000 invocations, depending on the LLM used.
Nonetheless, the approach faces several challenges: not all functions are suitable for translation, and for some, the amortization threshold is significantly higher or unreachable.
Despite these limitations, we identify four key research challenges whose resolution could unlock long-term, automated mitigation of energy debt in serverless computing.
\end{abstract}

\end{@twocolumnfalse}]
\section{Introduction} %

Cloud computing promises cost-effective virtually infinite on-demand resources, fueling a decades-long migration to the cloud. 
The increasing demand for cloud resources has consequently led to a predictable rise in energy consumption and global carbon emissions associated with the IT sector \cite{mastelic2015cloud,2024WorldBankITUmeasuring_emissions_energy}. 
To combat this worrying trend, cloud providers and application developers are being urged to adopt more energy-efficient practices and implement strategies to minimize their carbon footprint, in line with the United Nations sustainable development goals\cite{UN2015goals} and to consider sustainability as a software quality~\cite{2015LagoEtAlframing_sustainability_property}.
Although most cloud providers have already taken measures to lower their carbon emissions \cite{pedram2012efficientdatacenters}, a significant amount of emissions is caused by long-standing design choices in already running applications.

It is already well known that the selection of programming language~\cite{2017PereiraEtAlenergy_efficiency_programming}, the choice of architecture style~\cite{2025-Werner-ICSA-Clue}, and the choice of cloud technologies~\cite{lyu_myths_2023} can have a significant impact on the energy profile of an application.
While research and industry are developing new recommendations and new methods to reduce energy consumption for all these choices, most applications will not adopt them just for the sake of saving energy. 
Legacy applications, particularly those that are merely maintained and no longer actively developed, will remain long-lasting energy consumers operating on remote infrastructure. 
Hence, application owners have very little incentive to change their applications.
We call the cost of these long-past taken energy-consuming design decisions energy debt, a form of technical debt~\cite{2012_Kruchten_TechDept}.

While the cloud provides quick access to infinite cloud resources, it also makes it easy to forget about resources or lose track of them due to configuration drift or shadow IT~\cite{klotz2019causing}.
Hence, we argue that cloud providers are in a unique position to help application owners reduce energy debt, especially for legacy applications.
This can not only be beneficial to address the environmental sustainability responsibility of the cloud providers but also help them to deprecate old and inefficient platform technologies by moving applications away from them.

One unique cloud technology that can help in automatically reducing energy debt is serverless computing.
Serverless computing is a cloud computing model that abstracts away the underlying infrastructure, allowing developers to focus on writing code without worrying about server management.
This also enables cloud providers to propose and apply automatic improvements in the backend.
For example, Amazon Web Services (AWS) Lambda already tries to automatically upgrade the runtime of the serverless functions to the latest version, which can bring performance and security improvements.
However, we argue that serverless computing platforms can do even more, as they typically have access to the source code of the application.
This can allow the serverless provider to change the programming language, the architecture style of the application. 
Something that has already been explored to improve other serverless application qualities, for example, function fusion~\cite{2022_IC2E_Schirmer_Fusion} to reduce cold starts.

In this paper, we propose a new approach to reduce energy debt in serverless applications by leveraging the unique capabilities of serverless computing and large language models. 
Asking the research question: \textbf{How can we enable cloud computing platforms to automatically improve the energy efficiency of running (legacy) applications through automated code translation?}

For this, we make the following contributions:
\begin{enumerate}
    \item Evaluate the energy reduction potential in serverless applications by changing the programming language manually.
    \item Implement \Name{} an automatic serverless code transformation tool that can be integrated into existing platforms such as Fission.
    \item Evaluate the energy reduction potential of \Name{} using current large language models (LLMs).
    \item We identify four key challenges that must be addressed to realize \Name{}.
\end{enumerate}

The remainder of this paper is structured as follows: First, in \Cref{sec:rw} we review related work in the area of energy efficiency, serverless computing, and automatic code transformation.
Then, in \Cref{sec:problem} we describe the problem of energy debt and how it can be addressed, particularly in serverless computing, including a first evaluation of the energy reduction potential.
Following that, we introduce \name{} in \Cref{sec:design} and describe its integration into the existing serverless platforms Fission and evaluate \name{} using different configurations and LLMs in \Cref{sec:eval}.
Finally, we conclude the paper in \Cref{sec:conclusion} and discuss future work.

\section{Related Work} %

Addressing the environmental and technical sustainability of software systems is a major challenge for the software engineering community~\cite{2015LagoEtAlframing_sustainability_property}.
This is especially relevant in the context of cloud computing, where the cost and impact of engineering for sustainability are spread across multiple stakeholders, including cloud providers, application developers, platform engineers, and end-users~\cite{2025-Werner-ICSA-Clue}.
At the data center level, hyperscalers are already investing in renewable energy sources and custom hardware to improve energy efficiency~\cite{mastelic2015cloud,Chaudhari2023EstimatingPC,2019_Pierson_DATAZERO}.
Similarly, on the software engineering level, practitioners are investigating how various design patterns influence energy consumption, for example, within microservice architectural style~\cite{2025-Werner-ICSA-Clue, 2025-Bogner-ICSA-MSPatterns,centofanti2024impact}. 
More specifically, for software engineering, researchers are exploring the improvement of platform efficiencies~\cite{sharma2023challenges}, minimizing energy consumption and communication~\cite{fang2023energy}, enabling dynamic energy and emission controls~\cite{2020FieniEtAlsmartwatts_selfcalibrating_softwaredefined,2023_Thiede_SSoC_CarbonContainers,SilvadeSouza2020ContainergyACE} as well as evaluating the impact of auxiliary services~\cite{dinga2023empirical,boreges_observe_icsa_2024}.
This also includes the application level with a focus on the energy efficiency of the code itself.
Pereira et al.,~\cite{2021PereiraEtAlranking_programming_languages,2017PereiraEtAlenergy_efficiency_programming} evaluate the energy usage of several programming languages and show a significant difference of almost two orders of magnitude.
However, they observed that the relationship between energy consumption and factors such as time, memory, and resource varies across programming languages.
Hence, it is evident that programming languages can have a significant influence, though the extent depends on the program type and programming context. For instance, scientific computing applications often involve heavy computation, making language choice affects power usage. Additionally, the context in which a program is developed -- such as target hardware, runtime environment, or resource constraints -- can amplify or reduce this influence.
In this work, we are using this difference in programming language efficiency to enable data centers and platform operators to help application developers make more sustainable decisions.

Nevertheless, current cloud providers are still not equipping users with the necessary transparency and tools to reduce application-side energy consumption \cite{mytton2020assessing}, which in aggregate might reduce the emissions of data centers more than simply using renewable energy or buying sufficient carbon offsets.
In addition, even if developers are made aware of the energy consumption, they often do not act on it as this kind of re-engineering effort is often not rewarded~\cite{2023NoureddineEtAlimpact_green_feedback}.
Hence, with \name{} we aim to provide a tool that performs this task automatically, leaving the developer only with the task of reviewing and accepting the changes. Thus, reducing but not eliminating the development effort needed to pay off energy debt.

Automatically transforming code is a long-standing challenge in software engineering, with a variety of approaches proposed over the years. 
Most recently, the use of machine learning techniques, particularly large language models (LLMs), has gained traction in this area.
Here, emerging models such as CodeT5~\cite{2021WangEtAlcodet5_identifieraware_unified}, translation models~\cite{2022SzafraniecEtAlcode_translation_compiler}, and more generalized coding models such as Qwen-Coder~\cite{2025QwenEtAlqwen25_technical_report}, Gemma~\cite{2025TeamEtAlgemma_3_technical}, or OpenAI's GPT-4~\cite{openai2024gpt4technicalreport} have shown promising results in generating and transforming code.
Thus, researchers have already started to explore the potential of LLMs in enhancing the performance of systems~\cite{cui_large_2025}, as well as to improve the overall efficiency \cite{niu_evaluating_2024}.
In this work, we build on these recent advances in LLMs and propose a novel approach to automatically transform code for energy efficiency in the specific context of serverless computing.
\label{sec:rw}

\section{Cloud Energy Debt}\label{sec:problem} %

The energy consumption of a software system is strongly influenced by the architecture, programming language, execution runtime, hardware, and workload of the application~\cite{2015LagoEtAlframing_sustainability_property,2017PereiraEtAlenergy_efficiency_programming,2025-Werner-ICSA-Clue}.
For example, systems that employ machine learning models or integrate with proof-of-work blockchains inherently consume more energy. 
However, while these choices are often made knowingly and are often a requirement of the context and functionality of the application, the choice of programming language is not necessarily a design decision that is made due to context or functionality, but due to skill and team preferences and for long-term maintainability.
Consequently, design choices made years ago persist within applications, allowing energy inefficiencies to accumulate over time. This phenomenon parallels the concept of technical debt~\cite{2012_Kruchten_TechDept}, and we refer to it as energy debt.

\subsection{Towards Automated Code Translation}

Unlike technical debt, where eventually the worsening of maintainability will drive the team to refactor the code, energy inefficiency rarely motivates such action--unless it also impacts cost or performance.
Hence, we argue that we should address this energy debt through automation, e.g., automatic code refactoring to an energy-efficient language/runtime on the execution platform side.
However, many platforms, especially in the cloud, do not have access to the executable code. Hence, code refactoring options are limited.
Notably, the serverless paradigm is one of the few cloud models where the platform has access to the code and is fully in charge of the execution environment. 
Thus, we argue that serverless platforms have a unique opportunity to automatically repay the energy debt of legacy applications by refactoring the code to a more energy-efficient language/runtime.
Hence, in this work we focus on how we can design serverless platforms such that they can improve the energy efficiency of (legacy) functions through automated code translation as a first step towards answering our research question.

\subsection{Serverless Energy Savings Potential}
However, before we can answer the research question, we need to evaluate how much energy is saved by changing the language/runtime of a function.
For this, we selected Python as the source language, given its popularity in serverless function languages, and its relatively low energy efficiency~\cite{2017PereiraEtAlenergy_efficiency_programming}.
Similarly, we selected Go as the target language for translation, as it is still a very common serverless language with a comparatively high energy efficiency.
Based on Pereira et. al.~\cite{2021PereiraEtAlranking_programming_languages}, we can expect an up to 24x improvement. However, this included a broader range of workloads than typically found in serverless computing. 
\begin{table}
    \centering
    \caption{Functions used for the model evaluation.}
    \label{tab:functions}
    \resizebox{\columnwidth}{!}{
    
\begin{tabular}{llp{5.5cm}l}
\toprule
Function & Type & Description & Source\\
\midrule
$f_1$ & Basic & Basic ”hello world” function. & B \\
$f_2$ & Basic & Basic addition function.  & A  \\
$f_3$ & Basic & Basic calculator function with error handling.  & A \\
$f_4$ & Basic & Basic percentage calculator.  & A \\
$f_5$ & Basic & Compound interest calculator. & A  \\
$f_6$ & Basic & Recursive Fibonacci function.  & A \\
$f_7$ & Data Structures & Google Chat webhook function. & B \\
$f_{8}$ & Basic & Graph theory solver function. & C\\
$f_{9}$ & API & Performs HTTP GET requests to external API and returns transformed output. & D\\
$f_{10}$ & API & Fetches current weather data using OpenWeatherMap API. & D \\
$f_{11}$ & Data Structures & Extract and validate user information with nested data structures. & B\\
$f_{12}$ & Data Structures & Performs ELT on large JSON files. & A\\
$f_{13}$ & Syntactic Sugar & Uses a decorator pattern for logging. & D\\
$f_{14}$ & Syntactic Sugar & Implements a multi-stage data processing pipeline. & A\\
\bottomrule
\end{tabular}

    }
\end{table}

To get a more realistic picture of what we could expect, we collected a set of serverless functions (see \Cref{tab:functions}) that are written in Python and manually translated them to Go. 
We collected these functions from several sources: (1)~from simple introductory programming tasks, (2)~online examples, e.g., from AWS or Google's documentations, (3)~some we created based on CodeNet~\cite{puri2021codenetlargescaleaicode}, and (4)~some we created from API documentation.

\begin{figure}
    \centering
    \includegraphics[width=\columnwidth]{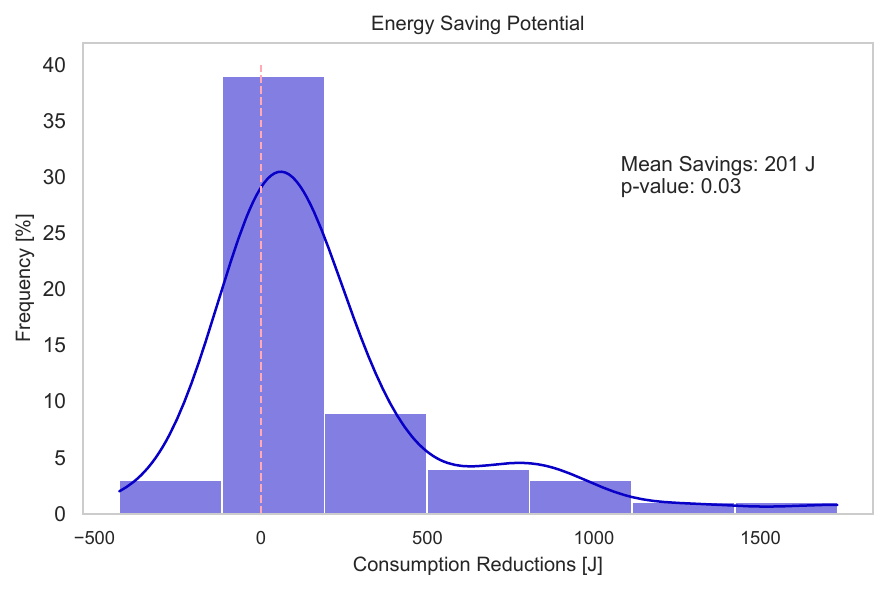}
    \caption{Energy savings per invocation, when manually translating the test functions (\Cref{tab:functions}) from Python to Go. Negative values indicate that the Python Version was more energy efficient.}
    \label{fig:saving_potential}
\end{figure}

We then run each function in a minimal container and invoked it \textbf{1,000} times using JSON-based events, measuring energy consumption in joules per invocation along with other performance indicators such as runtime and CPU usage(see Function-Metrics~\Cref{tab:metrics}).
Each function is running on a single core, without any parallelism in a closed-loop setting. 
Hence, consuming the JSON test events as fast as possible (see an example event in Listing~\ref{lst:input}).
We repeated these experiments five times and accumulated the results in~\Cref{fig:saving_potential}.
Note, we also collected the results of each of these invocations to later use as indicators for correct translation of functions.

Based on these first experiments, we can see that the potential for savings exists. 
The savings are minimal for most of the tested functions, as the functions were rather basic.
However, we already see larger savings for the more complex functions, such as the \textit{data structures}. 
Moreover, even if the savings are small in absolute joules, they already represent a 70\% average reduction in consumption.

However, we also see that some functions present with an increased consumption. 
Notably, these were the basic functions that did simple calculations. 
We assume that Python has a more efficient way to deserialize and serialize the JSON events, as this is the majority of the execution time used for these functions. 

Thus, the savings depend on the function context, creating a varying translation budget for each function that we can spend to save energy after a reasonable number of invocations. 
This indicates that we need a way to predict if functions have a chance of reducing energy before attempting automatic translation, which is one of the key future challenges (\textbf{C1}) that we identified.
Nevertheless, we see the potential, especially as more complex functions all showed strong saving potential and thus present a first system design in the following section that can perform automated code translation.

\section{\Name{}}\label{sec:design} %

In the following, we describe the design of \name{} (\NameLong{}) and how we integrated it into the Fission\footnote{\url{https://fission.io/docs/architecture/}} serverless platform, as an example.

\subsection{Design of \name{}}
\Name{} is designed as a standalone service that expects a deployment package (e.g., a zip file) containing the function code and configurations to build.
In addition, \name{} also requires a set of test files (see Listing~\ref{lst:befor},\ref{lst:after}), which are used to evaluate the function after it has been translated.
Each test file contains an input event (as JSON) and an expected output response (as JSON). 
\Name{} treats both the deployment package and the resulting artifact as black boxes; hence, the tests conducted as black box tests.
Serverless computing comes with fixed invocation interfaces, which makes it rather simple to run these black-box tests against implementations of different languages.

\begin{figure}
    \includegraphics[width=\columnwidth]{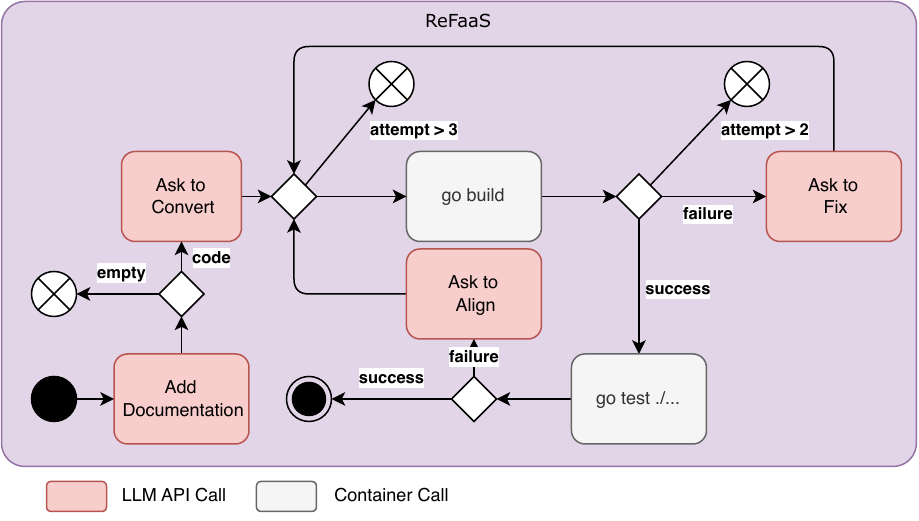}
        \caption{Translation Pipeline used in \name{}, highlighting the Chain of Thought approach with external validation (gray boxes) that is used to achieve high translation quality.}\label{fig:pipeline}
        \vspace{2em}
\end{figure}

\Name{} enables the use of several different LLMs to translate functions.
Moreover, we can also use different LLMs at each stage of the translation process. 
For this, \name{} provides an extensible pipeline interface that allows defining the LLM-Client, prompt, and verification and recovery steps. The pipeline is defined in a YAML file, which allows for defining the LLMs and their parameters. 
The verification steps are used as gateways to stop faulty or unnecessary translations.
For example, if a function does not compile, we can stop trying to test it. 
Moreover, we allow for multiple recovery steps, for example, prompting the LLM to fix the code given the compilation error message and running the compiler again until we get to a buildable version. 
The main pipeline we used is presented in \Cref{fig:pipeline}. 

For the pipeline, we follow a multi-shot, Chain-of-Thought approach. The pipeline consists of three main steps: (i) the code translation, (ii) the building of the new artifacts, and (iii) the testing. 
In each step, we also added additional steps to solve problems (recovery) to increase the likelihood of a successful translation.
For example, in the code translation step, we added a step to ask the LLM to first document the original code excessively.
In case the build step fails, we also have a step to ask the LLM to resolve the build errors.
Lastly, in the testing step, we ask the LLM to ensure that the code is aligned with the original by providing the original source code and the current source code as input to the LLM. 
Each step, including the recovery steps, is always limited to a fixed number of attempts to ensure termination.

\Name{} operates a simple job queue, so that multiple conversion requests can be processed in parallel. 
However, the limiting resource is the used LLM API and underlying hardware. 
We specifically designed \name{} to use locally deployed LLMs, as the source code would typically be confidential, although commercial cloud-based LLMs can also be used, in less critical cases.
Moreover, \name{} also allows each of the steps to use a different LLM API, so that we can also combine more narrow LLMs, e.g., coding assistance LLMs, for some tasks.

Once a job is terminated, either successfully or unsuccessfully, a new deployment package is generated, either the original or the translated one. 
Hence, from the perspective of the serverless platform, it always gets a deployment package as a user would provide it.
\Name{} is available on GitHub at\anonymize{:~\url{https://github.com/ISE-TU-Berlin/ReFaaS}}.

\subsection{Fission Integration}
\begin{figure}
    \includegraphics[width=\columnwidth]{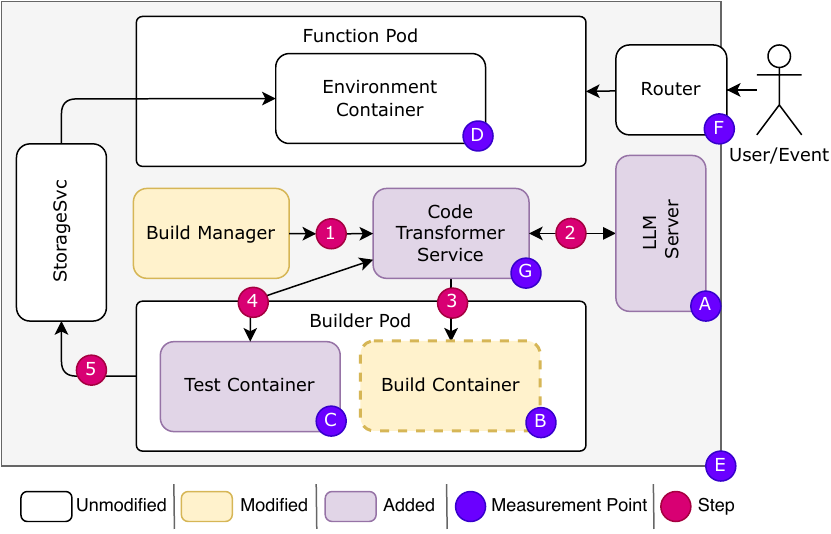}
    \caption{Integration of \name{} into Fission's architecture and process, including relevant Steps, changed components, and Measurement Points relevant for the evaluation in \Cref{sec:eval}).}\label{fig:fission_llm}
\end{figure}
\Cref{fig:fission_llm} shows how we can integrate \name{} into the Fission deployment process.
The Fission deployment process includes a build pod and an execution pod. 
The build pod is responsible for building the function so that it can be executed and scaled. 
The compiled and prepared build artifact is stored in a cluster-wide storage service. 
Hence, for us to integrate \name{}, we mainly need to modify the behavior of the build pod, and if a translation is successful, instruct the function pod to deploy a different environment container, i.e., a Go-based one instead of a Python environment.
We augmented the build process by adding a fork into the build manager, which first calls \name{} to translate the function \circled{1}. Once \name{} receives the request, it starts the conversion pipeline (\Cref{fig:pipeline}), which requires one or multiple API calls to the LLM Service \circled{2}.
Once we receive the translated source code, we build it using the build pod \circled{3}. 
Once the build is successful, we create a new test container \circled{4} and run the tests using the translated artifact.
If these tests conclude successfully, we store the translated artifact in the cluster-wide storage \circled{5} and update the function pod configuration to use the translated artifact.
Otherwise, we build the original artifact and store it in the cluster-wide storage \circled{5}.
Since \name{} works asynchronously, we can also first build the original artifact and deploy it while attempting to translate the function, thus minimizing potential cold-start delays.
If the translation is successful, we can roll out the change.

\subsection{Critical Assumptions}
For \name{} to be effective, we make several assumptions.
For one, we assume that the user provides a sufficient number of test events, e.g., archiving at least full test coverage in the original function. 
This gets increasingly difficult, and also less realistic for functions that have side effects, e.g., writing to a database or consuming an external API.
We've experimented with the use of Local-Stack, a set of mock-up services for AWS, as one option to offer a predictable environment for \name{} to run dynamic black box tests.
However, for now, we assume that the user is providing the test events and an appropriate test environment.
Moreover, we assume that the function is translatable, i.e., that there are no critical dependencies in the original that don't exist in the target language, e.g., something like TensorFlow.
We also assume the original function is correct and valid.

Moreover, \name{} currently performs semantics comparison of the resulting JSON responses (see Listing~\ref{lst:befor},\ref{lst:after}) to validate the test cases.
Here, small differences in the content of error messages or changes due to external influence (e.g., time) can change the semantic similarity between expected and test output. 
Hence, the current prototype assumes that the tests produce deterministic output. 
This is also an issue, in case the functions use external APIs without a means to mock an environment.
In general, the automatic translation validation is one of the future key challenges (\textbf{C2}) we identified for automatic code translation.

During the translation process, we measure the runtime and energy consumption of the translated function. 
Thus, we also try to detect if the translated function is consuming more resources than the original function for the same inputs, in which case we abort the translation process.
However, this still can incur energy, time costs that do not contribute to the goal \name{}.

\section{Evaluation}\label{sec:eval} %

In this section, we evaluate the performance of \Name{} in terms of translation success and the savings it can provide.
For this, we selected four state-of-the-art LLMs that can be run locally (for accurate energy measurements) and three pipeline configurations.
\begin{table}
    \centering
    \caption{Relevant metrics we collected in \name{} evaluation. }\label{tab:metrics}
    \resizebox{\columnwidth}{!}{
    \begin{tabular}{l|p{0.8\columnwidth}p{4em}}
    \textbf{Metric} & \textbf{Description} & Measurement Point\\
\hline
    \multicolumn{3}{c}{Conversion Metrics} \\
\hline
     Runtime [s] & Time it took to complete the pipeline for the function. & \measured{G} \\
     Token [\#] & Token it took the LLM to prompt and generate the response. & \measured{A} \\
     Energy [Wh] & The total energy consumption of \name{} and the LLM-Service for producing the code. & \measured{A},\measured{E} \\
     Temperature [\textdegree{}h] & The total accumulated temperature per hour that the server produced in additional heat during the code transformation. & \measured{A},\measured{E}\\
\hline
    \multicolumn{3}{c}{Function Metrics} \\
\hline
     Validation [binary] & Indicates if all black-box tests produced valid results. & \measured{C}\\
     Test [\#] & Number of tests that produced valid results. & \measured{C}\\
     Buildable [binary] & Indicates if the function was built without errors. & \measured{B}\\
     $\Delta$ Energy [Wh] & Change in energy consumption between the original and transformed function. & \measured{C}, \measured{D}\\
     $\Delta$ CPU [s] & Change in CPU consumption between the original and transformed function. & \measured{C},\measured{D} \\
     $\Delta$ Memory [MB] & Change in Memory consumption between the original and transformed function. & \measured{C},\measured{D}\\
     $\Delta$ ColdStart [s] & Change in cold start duration between the original and transformed function. & \measured{F}\\
\end{tabular}

    }
\end{table}
\subsection{Experiment Design}
Towards that end, we selected four models: \texttt{qwq} the reasoning model and \texttt{Qwen2.5-Coder} from \cite{2025QwenEtAlqwen25_technical_report}, \texttt{Deepseek-R1}~\cite{2025DeepSeek-AIEtAldeepseekr1_incentivizing_reasoning} a recent well performing reasoning model and \texttt{Gemma3}~\cite{2025TeamEtAlgemma_3_technical}.
For each model we tested three different pipeline configurations: (i) \textbf{Chain of Thought} (CoT), as described in \Cref{fig:pipeline} and \Cref{sec:design}, (ii) 
\textbf{Creative CoT}, the same as CoT but with a high temperature of 0.8 instead of 0.1, and (iii) \textbf{Single-Shot}, where we followed the same pipeline as in CoT but without any repetitions or recovery steps.
We run each experiment three times and always utilize all test functions (see \Cref{tab:functions}) in random order.
In all cases, we performed translations from Python to Go functions. 
We mainly selected Go over other, more energy-efficient languages such as C++, Rust, or C, due to the authors’ familiarity with Go, which enabled manual evaluation of translation accuracy.
During the experiments, we measured the time and energy consumption of the translation process, as well as the time and energy consumption of the resulting functions. 
A detailed description of the metrics can be found in \Cref{tab:metrics}, for example, to measure the conversion Energy $[Wh]$, we deployed Scaphandre\footnote{\url{https://github.com/hubblo-org/scaphandre}} to probe the energy consumption of each container in the Fission namespace \measured{E}, and the dcgm-exporter\footnote{\url{https://github.com/NVIDIA/dcgm-exporter}} to collect the consumption of the GPU \measured{A} in \Cref{fig:fission_llm}.
We also measured the accumulated temperature of the GPU and CPU using dcgm-exporter and Prometheus node-exporter, as cooling significantly contributes to the energy consumption of the compute infrastructure~\cite{2024_Alkrush_Refrigeration_2024}.
We performed all experiments on a Precision 3680, using an Nvidia 4500 Ada GPU, an Intel i7 14700K, and 64 GB of memory.

For the stability of \Name{} and the selected pipeline, we observed (i) if a function was buildable after finishing the pipeline ($\text{Buildable}$), (ii) how many of the tests passed after a translation ($\text{Test}$) and (iii) if all tests of the transformed function passed ($\text{Validation}$).
If a function was valid, we ran the same micro-benchmark as described in \Cref{sec:problem}.

\subsection{Experiment Results}
\begin{table*}
    \caption{Aggregated experiment results that resulted in at least 25\% of buildable functions. }\label{tab:headline_results}
    \resizebox{\textwidth}{!}{
    \begin{tabular}{llp{1.8cm}p{1.8cm}p{1.8cm}p{1.8cm}p{1.8cm}p{1.8cm}p{1.8cm}}
\toprule
 &  & Validated [\%]$\uparrow$ & Tests [\%]$\uparrow$ & Buildable [\%]$\uparrow$ & $\overline{\text{Runtime}}$ [s]$\downarrow$$^*$ & $\overline{\text{Tokens}}$  [\#]$\downarrow$$^*$ & $\overline{\text{Energy}}$ [Wh]$\downarrow$$^*$ & $\overline{\text{Temp.}}$ [\textdegree{}h]$\downarrow$$^*$ \\
Model & Configuration &  &  &  &  &  &  &  \\
\midrule
\multirow[t]{3}{*}{qwq} & Chain of Thought & 36 & 35 & 50 & 177 & 4685 & 14 & 7 \\
 & Creative-CoT & 21 & 24 & 50 & 45 & 4187 & 7 & 3 \\
 & Single-Shot & 14 & 20 & 64 & 22 & 2448 & 5 & 2 \\
\cline{1-9}
\multirow[t]{3}{*}{qwen2.5-coder 32b} & Creative-CoT & 43 & 51 & 79 & 41 & 2319 & 7 & 3 \\
 & Chain of Thought & 36 & 39 & 71 & 127 & 2534 & 11 & 6 \\
 & Single-Shot & 29 & 39 & 71 & 19 & 1462 & 5 & 2 \\
\cline{1-9}
\multirow[t]{2}{*}{gemma3 27b} & Single-Shot & 43 & 41 & 43 & 28 & 2075 & 5 & 3 \\
 & Creative-CoT & 36 & 39 & 50 & 51 & 3366 & 7 & 3 \\
\cline{1-9}
deepseek-r1 32b & Chain of Thought & 14 & 12 & 29 & 37 & 4386 & 6 & 3 \\
\cline{1-9}
\bottomrule
\end{tabular}

    }
     \vspace{0.9em}$^*$~Only successful translations are considered.
\end{table*}
\Cref{tab:headline_results} shows the overall results of the evaluation with a buildable ratio of at least 25 percent, showing that the better-performing pipelines produce up to 79 percent of buildable functions.
Additionally, the Creative-CoT \texttt{Qwen2.5-Coder} and Single-Shot \texttt{Gemma3} yield the best results with a 43 percent translation success rate.
Notice that \texttt{Deepseek-R1} did not perform well, which is mainly due to extraction problems, where the model would always return more text than the requested code, causing issues in parsing and using the output.
For validation of the functions, we use the same JSON-based test files in both Python and Go and then compare the output semantically. For each function, we used between 4 test inputs. We selected the test inputs to achieve a high degree of code coverage and test important edge cases. 
\begin{figure}
\centering
\begin{subfigure}{0.9\columnwidth}
        \centering
\begin{lstlisting}[language=ieeepython, caption={Original Python Function for $f_{5}$},label={lst:original}]
#...
try:
    result = num1 / num2
    return success(result)
except Exception as e:
    return error(400, str(e))
#...
\end{lstlisting}
        \end{subfigure}
        \begin{subfigure}{0.9\columnwidth}
            \centering
\begin{lstlisting}[language=ieeego, caption={Generated Go function for $f_{5}$},label={lst:translated}]
//...
if num2 == 0 {
	return error(400, "division by zero")
} else {
	return success(num1 / num2)
}
//...
\end{lstlisting}
        \end{subfigure}
        \begin{subfigure}{0.9\columnwidth}
            \centering
\begin{lstlisting}[language=ieeejson, caption={Test2 input for $f_{5}$},label={lst:input}]
{
    "num1": 1, "num2": 0, "operation": "/"
}
\end{lstlisting}
        \end{subfigure}
        \begin{subfigure}{0.9\columnwidth}
            \centering
\begin{lstlisting}[language=ieeejson, caption={Expected Output of Test 2 in $f_{5}$},label={lst:befor}]
{
    "statusCode": 400,
    "body": "{\"error\": \"ZeroDivisionError: division by zero\"}"
}
\end{lstlisting}
        \end{subfigure}
        \begin{subfigure}{0.9\columnwidth}
            \centering
\begin{lstlisting}[language=ieeejson, caption={Output for Test 2 using Listing~\ref{lst:translated}},label={lst:after}]
{
    "statusCode": 400,
    "body": "{\"error\": \"division by zero\"}"
}
\end{lstlisting}
        \end{subfigure}
        \caption{Example of a translation error.}\label{fig:issue}
\end{figure}
Here, we often saw issues, especially for error-handling cases, see \Cref{fig:issue}. 
In the original functions (Listing~\ref{lst:original}), some errors were just JSON-wrapped Python errors, which, even if translated faithfully (as in Listing~\ref{lst:translated}), would not result in the same type of output (Listing~\ref{lst:befor},\ref{lst:after}). 
This highlights one of the main challenges in translating functions, which is a way to fairly and accurately validate the generated code (\textbf{C2}).

\begin{figure}
    \centering
    \includegraphics[width=.9\columnwidth]{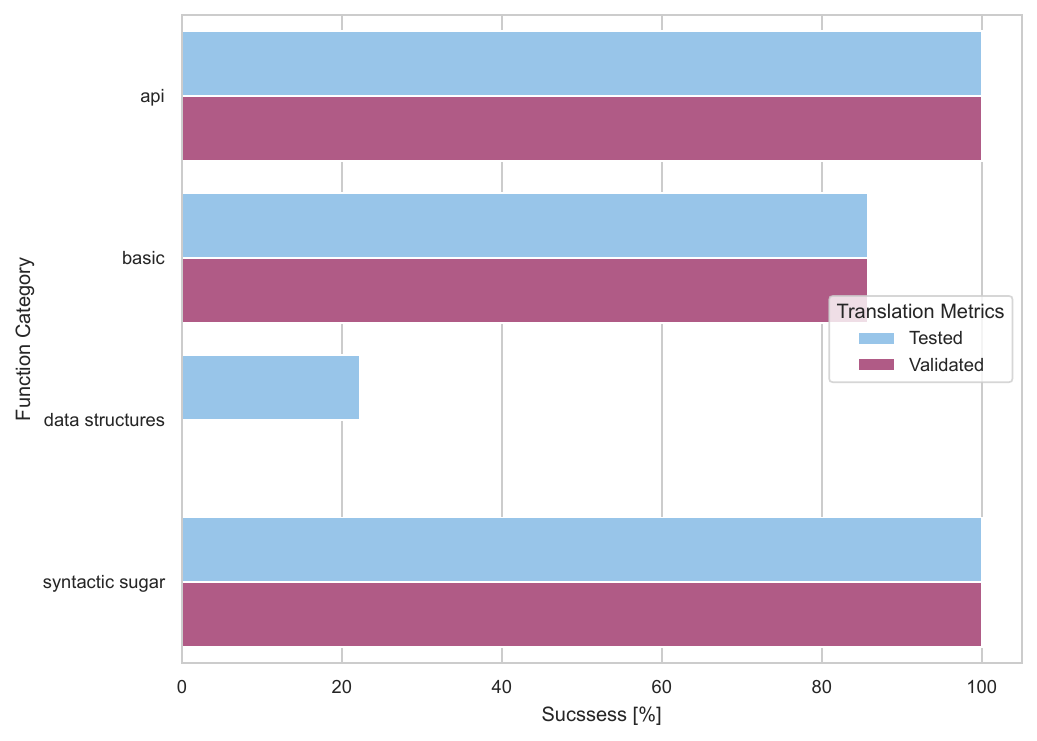}
   \caption{Aggregated code translation success across all models and configurations.}\label{fig:translation}
\end{figure}

Still, we see that the models can translate a function into a valid form, as we have seen in the validation results.
However, only 43 percent is not satisfying, as this implies a large number of functions are not yet translatable into a valid version.
However, if we look across all runs, we can see that most functions were translated in a valid form at least once, see~\Cref{fig:translation}.

We assume that the size of the used models, or the number of recovery attempts, was not sufficient to reach a valid translation.
Similarly, it could be that more test cases would have yielded better results by providing more errors for the LLM to align the solution more with expected outputs. However, here we have to also balance overfitting to test cases.
Thus, more thorough prompt engineering and use of larger models could improve the results.
However, the potential of the approach is evident from the validation percentage.

\begin{figure}
    \centering
    \includegraphics[width=0.95\columnwidth]{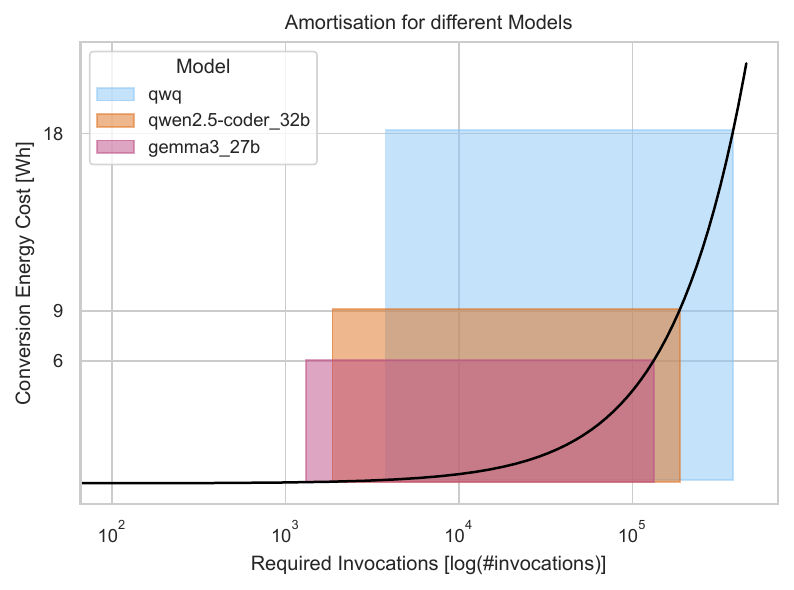}
    \caption{Number of invocations before the conversion is amortized. The area indicates the range between the best and worst case we observed within the test set \cref{tab:functions}.}\label{fig:function_amortisation}
\end{figure}
In terms of the cost associated with the valid translation of a function, we observe that these models range from 5 to 14 Wh, with an increased temperature load of 2 to 7 degrees per hour.
With a mean runtime of $~61$ seconds and $~3050$ tokens, which would amount to around 2 USD cents per function using the Google Gemini API.
\Cref{fig:function_amortisation} shows the savings potential for the three best-performing models and configurations in terms of invocations until the translation is amortized, i.e., the point at which the accumulated energy reduction of invoking the Go function is greater than the total energy consumption of using \name{}. 
The shaded area indicates the best and worst case for the given model, depending on how much energy reduction the function can achieve (see \Cref{fig:saving_potential}).
Here, we can see that some functions, with very little savings potential, require nearly $10^6$ invocations, which might be more than we can expect during the lifetime of some functions.
On the other hand, we can see that some functions would already be amortized after around 3000 to 5000 invocations.
\Cref{fig:function_amortisation} does not include failed attempts or functions that consume more energy than before the translation, as these costs can never be amortized. 
Instead, these costs would act as additional conversion costs, increasing the time until functions amortize.
Hence, to make the \name{}-approach viable, we need to reduce these two drivers, the energy consumption of failed attempts and the energy consumption of running \name{}.
The first driver could be managed by addressing challenge (\textbf{C1}) -- predict if functions have a chance of reducing energy before attempting automatic translation.
The second driver could be managed by finding more energy-efficient code translation models, another key future challenge (\textbf{C3}) we identified.

Overall, we can see, however, that we can design serverless platforms in such a way that they can automatically improve the energy efficiency of functions with the help of current large language models.
A more thorough investigation, using also different programming languages and a more robust set of test functions, could further refine our results (\textbf{C4}); however, currently no representative sample of real-world serverless functions exists to see how well \name{} would perform in practice.

\section{Conclusion}\label{sec:conclusion} %

In this paper, we identified the problem of energy debt, a special form of technical debt, that accumulates due to energy inefficient design choices.
We argued that developers are not incentivized or aware to pay off this debt, even less so than regular technical debt.
Hence, we propose \name{} an automatic code translation service that can be used to translate serverless functions from one language to another, which is one strategy to pay off energy debt.

\subsection{Discussion}
With \name{} we showcase the potential of automatic code translation to reduce the energy consumption of cloud services.
However, for this, the platform providers need access to the source code and configurations of the applications.
While this is not always the case, we argue that serverless platforms, such as Fission, are uniquely able to perform this and other automatic improvements. 

We showed that \name{} can translate serverless functions with a success rate of up to 43 percent, which is already a promising result, and that in the best case, such translation can save up to 70\% of invocation energy costs.
However, we also see several limitations with the current approach. 
Firstly, the selected functions were still basic, and we still need to compare \name{} with more complex functions. 
However, early coding benchmarks~\cite{2023JiaoEtAlevaluation_neural_code} already show promising trends even for complex functions, e.g., with dependencies.
Secondly, we used representative test inputs to validate the functions and also performed manual reviews of the generated code. 
However, with increased function complexity, this becomes increasingly difficult, requiring more sophisticated validation methods.
Here, we could see a human-in-the-loop design, where instead of automatically applying the refactoring changes, developers can review the changes together with potential energy and cost savings. 
While this is not a trivial task for developers it could reduce the barrier to getting started. 
Moreover, a switch from Python to Go can also reduce execution time and memory use, which translates to direct cost savings in the serverless model, hence, further incentivizing developers to take a look.
Thirdly, LLMs are prone to hallucinations, which also applies to generated code, creating an issue in providing stable translation time and code quality.
Here, we argue that the pipeline approach with multiple validation checkpoints (see \Cref{fig:pipeline}) can offer a strategy to improve translation stability.

\subsection{Challenges and Future Work}
Throughout the paper, we identified current limitations on using the \name{} approach. 
From them, we thus identified four main challenges that need to be addressed in future work:
\begin{description}
    \item[\textbf{C1}] \textbf{Translation Prediction} -- that can predict the likelihood that a function can be translated and will offer a reasonable energy reduction.
    \item[\textbf{C2}] \textbf{Automatic Translation Validation} -- programmatic means to fairly compare the behavior of two function implementations beyond back-box testing.
    \item[\textbf{C3}] \textbf{Energy-Efficient Translation Models} -- that can deliver good code-to-code transformations with less energy consumption.
    \item[\textbf{C4}] \textbf{Robust FunctionSet} -- that can offer a robust and broader set of functions to test the translation process, not only for the aim of energy reduction, but also for serverless evaluation in general.
\end{description}

Nevertheless, we believe that these challenges can and will be addressed in the future, as already promising work is being done in these areas.
Moreover, while \name{} only performed code translation, other automatic refactoring could follow a similar approach, e.g., changing configurations or replacing known energy-consuming libraries. 
Thus, offering further options to code once and run green.

\bibliographystyle{plain}
\bibliography{main}

\end{document}